\documentclass[a4paper,11pt]{article}
\usepackage[latin1]{inputenc}
\usepackage{amssymb}
\usepackage{amsmath}
\usepackage{graphicx}
\usepackage{epsfig}
\usepackage{enumerate,epsfig}
\usepackage{subfigure}

\usepackage{pstricks, pst-node, pst-tree}
\usepackage{hyperref} 

\newtheorem{Def}{Definition}
\newtheorem{Bsp}{Example}
\newtheorem{Bem}{Remark}
\newtheorem{Satz}{Proposition}
\newtheorem{HSatz}{Theorem}

\newenvironment{Ack}{\par {\bf Acknowledgment.}}{}

\newcommand{\setN}{\mathbb{N}} 
\newcommand{\setZ}{\mathbb{Z}} 
\newcommand{\setC}{\mathbb{C}} 
\newcommand{\kC}{\mathcal{C}}	
\newcommand{\kD}{\mathcal{D}}	
\newcommand{\kH}{\mathcal{H}}	
\newcommand{\kJ}{\mathcal{J}}	
\newcommand{\kE}{\mathcal{E}}	
\newcommand{\kF}{\mathcal{F}}	
\newcommand{\kG}{\mathcal{G}}	
\newcommand{\kK}{\mathcal{K}}	
\newcommand{\kS}{\mathcal{S}}	
\newcommand{\kP}{\mathcal{P}}	
\newcommand{\kX}{\mathcal{X}}	
\newcommand{\kY}{\mathcal{Y}}	
\newenvironment{Proof}{ {\it Proof.}}
							{\hspace*{\fill} $\Box$\par\vskip2ex}	

\title{Equality conditions for internal entropies of certain classical and 
quantum models}
\author{Peter Gmeiner\footnote{Department Mathematik, Friedrich-Alexander-Universit\"at Erlangen-N\"urnberg, Bismarckstra{\ss}e 1 1/2, D-91054 Erlangen, Germany. E-mail: gmeiner@mi.uni-erlangen.de}}
\date{\today}

\begin{document}

\maketitle

\begin{abstract}
Mathematical models use information from past observations to generate 
predictions about the future. If two models make identical predictions the one 
that needs less information from the past to do this is preferred. It is 
already known that certain classical models (certain Hidden Markov Models 
called $\epsilon$-machines which are often optimal classical models) are not in
 general the preferred ones. We extend this result and show that even optimal 
classical models (models with minimal internal entropy) in general are not the 
best possible models (called ideal models). Instead of optimal classical 
models we can construct quantum models which are significantly better but not 
yet the best possible ones (i.e. they have a strictly smaller internal 
entropy). In this paper we show conditions when the internal entropies between 
classical models and specific quantum models coincide. Furthermore it turns 
out that this situation appears very rarely. An example shows that our results 
hold only for the specific quantum model construction and in general not for 
alternative constructions. Furthermore another example shows that classical 
models with minimal internal entropy need not to be related to quantum models 
with minimal internal entropy.
\end{abstract}

\section{Introduction}

Mathematical modeling of natural and technological systems plays an important 
role in modern science. In general, there are many ways to model a system 
mathematically. One possibility is to view the system of interest as an 
information processing black box generating an observable output from given 
past observations. The observed data can be treated as a stochastic process 
and we try to find models which are called {\em Hidden Markov Models}, that 
generate the same statistical behaviour and that are denoted as classical 
models. We prefer models which predict future data from past observations in 
an optimal way, i.e. they need as little memory as possible to do this. The 
amount of information the past contains about the future is measured by the 
mutual information between past and future data. This quantity is known as 
{\em excess entropy} \cite{Cru83}. A model that should be able to predict 
future data in an optimal way has at least to store this amount of information 
to do this. One method to construct such a model in a systematic way is used 
in computational mechanics and called (classical) {\em $\epsilon$-machine}. 
$\epsilon$-machines are the optimal classical models for a certain subset in 
the set of all possible alternative Hidden Markov Models but not the optimal 
classical models in general. The optimality of a classical model is quantified 
by the classical internal state entropy of the model. Usually this is the 
Shannon entropy and for an optimal classical model the internal state entropy 
is called {\em generative complexity} $C_{Cl}$. Instead of considering 
classical models one can think about analog quantum models (called 
{\em Hidden Quantum Markov Models}). Recent results show that if the classical 
$\epsilon$-machine is not already the best possible model (called {\em ideal 
model}), it is always possible to find a quantum model that needs less memory 
than the classical $\epsilon$-machine to reconstruct the statistical behaviour 
of the stochastic process \cite{GuW11}. Usually the internal state entropy 
$C_q$ of the quantum model is strictly greater than the excess entropy $E$ and 
there remains room for improvement. We extend this results for all optimal 
classical models. 

The Hidden Quantum Markov Model induced from a classical Hidden Markov Model, 
can be formulated in the setting of a {\em quantum channel}. The initial 
distribution and the transition probabilities of a classical Hidden Markov 
Model (Definition \ref{DefHMM}) can be used to calculate the mutual 
information $I(X;Y)$ between a specific classical input random variables $X$ 
and a classical output random variables $Y$ related to the classical model. We 
achieve the following inequality chain in the subsequent sections
$$E \le I(X;Y) \le C_q \le C_{Cl}.$$
In this paper we investigate for a specific quantum model construction 
equality conditions for the last two inequalities above. We will see that in 
general there remains a gap between the different internal state entropies for 
the suggested quantum model construction introduced in \cite{GuW11} and that 
the last two inequalities are strict in most cases. Furthermore for 
$\epsilon$-machines we prove that $E=I(X;Y)$ hold and show with an example 
that a quantum model induced by a minimal classical model is not the minimal 
quantum model. The relationship between minimal classical models and minimal 
quantum models remains an open question.

This paper is organized as follows. In Section \ref{sec:pre} some basic 
notations and definitions are introduced. Section \ref{sec:minhmm} introduces 
$\epsilon$-machines, restates a recently proved theorem and extends this 
theorem to minimal Hidden Markov Models. Section \ref{sec:qua} introduces 
Hidden Quantum Markov Models. Furthermore two well-known propositions applied 
to our context are presented and we generalize a further theorem from 
$\epsilon$-machines to minimal Hidden Markov Models. The example which shows 
that minimal classical models do not correspond to minimal quantum models is 
also presented here. In Section \ref{sect:equcond} we prove the equality 
conditions for the internal entropies and in Section \ref{sec:ex} we present a 
calculation example and verify the proven results. Section 
\ref{sec:alternatives} describes an alternative construction of a quantum 
model to model a stochastic process and shows that the equality conditions in 
Section \ref{sect:equcond} in general cannot be extended to other quantum 
model constructions than the suggested one in Section \ref{sec:qua}.

\section{Preliminaries} \label{sec:pre}

Let $(\Omega, \kF, P)$ be a probability space with a metric space $\Omega$, a 
$\sigma$-algebra $\kF$ and a probability measure $P$. 
For random variables $X, Y: \Omega \rightarrow \Sigma$ mapping to a finite 
alphabet $\Sigma$ the Shannon entropy is defined by
$$
H(X) := - \sum_{x \in \Sigma} P(X = x) \log P(X = x),
$$
and the conditioned Shannon entropy by
$$
H(X | Y) := - \sum_{x, y \in \Sigma} P(X = x, Y = y) \log P(X = x | Y = y),
$$
where 
$P(X=x) := P\left ( \left \{\omega \in \Omega | X(\omega) = x \right \} \right)$
 denotes the probability that the random variable $X$ is equal to 
$x \in \Sigma$, $P(X=x, Y=y)$ is the joint probability between $X$ and $Y$ and 
for $P(Y=y) > 0$ the conditional probability is 
$P(X=x|Y=y) := \frac{P(X=x, Y=y)}{P(Y=y)}$.
In the definitions the convention $0 \log(0) = 0$ is used. Given a distribution
 $\mu$ of a random variable $X$ we sometimes write $H(\mu)$ instead of $H(X)$. 
The mutual information between two random variables is 
$$
I(X;Y) := H(X) - H(X|Y).
$$
The mutual information is non negative $(I(X;Y) \ge 0)$ and equals zero if and 
only if $X$ and $Y$ are independent random variables \cite{Cov06}.

We consider a time-discrete stationary stochastic process 
$\overleftrightarrow{X} := (X_t)_{t \in \setZ}$ with random variables 
$X_t : \Omega \rightarrow \Sigma$ for all $t \in \setZ$. We define the 
semi-infinite processes $\overleftarrow{X} := (X_{-t})_{t \in \setN}$ 
interpreted as past and $\overrightarrow{X} := (X_t)_{t \in \setN_0}$ 
interpreted as future respectively. Blocks of random variables with finite 
length are denoted by $X_a^b := (X_k)_{k\in [a,b]\cap \setZ}$ for 
$-\infty < a \le b < \infty$. The one-sided sequence space is 
$\Sigma^\setN := \times_{i \in \setN} \Sigma$ and in the same way the two-sided
 sequence space $\Sigma^\setZ$ is defined. We introduce the shift function 
$\sigma: \Sigma^\setZ \rightarrow \Sigma^\setZ$ by $\sigma(x)_i := x_{i+1}$.
At any time $t \in \setZ$ we have random variables 
$X_{-\infty}^t := (X_k)_{k \le t}$ and $X_{t+1}^\infty:=(X_k)_{k \ge t+1}$ that
 govern the systems observed behaviour respectively in the shifted past and the
 shifted future. The mutual information between these two variables is the 
well-known {\em excess entropy} \cite{Cru83, Cru03}
\begin{equation} \label{ExcessDef}
E := \lim_{L \rightarrow \infty} I(X_0^{L-1} ; X_{-L}^{-1}).
\end{equation}
In general, it is not clear if the limit in (\ref{ExcessDef}) exists. We will 
see later that in the setting of this paper $E$ always exists. With the 
assumption that the limit in (\ref{ExcessDef}) exists as a finite number the 
following equality holds: $E = I(\overleftarrow{X}; \overrightarrow{X})$, see 
Chapter 2.2 in \cite{Pin64}.

The stochastic process generates a sequence of output symbols which represents 
the observed behaviour of a system for which we construct a mathematical model 
in a discretized fashion.

We use a Hidden Markov Model (HMM) to model a given stochastic process. In 
general there are different kinds of HMMs. For our purpose we use a 
transition-emitting HMM and use the same terminology as in 
\cite{Loe10, Loe09a, Loe09b}.

\begin{Def} \label{DefHMM}
With $\kP(A)$ we denote the space of all probability measures on a set $A$. A 
{\bf transition-emitting HMM} consists of a set $\kS$ of {\bf internal states} 
and a pair $(T,\mu)$ with an initial distribution $\mu \in \kP(\kS)$ and a 
measurable function $T : \kS \rightarrow \kP(\kS \times \Sigma)$, called 
{\bf generator}. We say that $(T,\mu)$ is an HMM of $\overleftrightarrow{X}$ if
 the output-distribution which is determined by the output kernel 
$K_s(.):= T(s)(\kS \times .)$, $s \in \kS$ of the HMM coincide with the 
distribution of $\overleftrightarrow{X}$.
\end{Def}

In the following we abbreviate transition-emitting HMM with HMM. Since we are 
considering stationary stochastic processes we require that the HMM is 
invariant in the following sense.

\begin{Def}
A HMM $(T,\mu)$ is {\bf invariant}, if $\mu$ is $T$-invariant i.e.
$$\mu(G) = 
		\int_{\kS} T(s)(G \times \Sigma) d\mu(s), \qquad \forall \, G \in \kS.
$$
\end{Def}

We are interested in HMMs with minimal internal state entropy $H(\mu)$ which 
can be considered as a complexity measure of the process generated by the HMM. 
Following L\"ohr \cite{Loe09c, Loe10} we define the generative complexity.

\begin{Def} \label{gencomplexity}
The (classical) {\bf generative complexity} of a stationary stochastic process 
$\overleftrightarrow{X}$ is the infimum of the entropies of internal states
$$C_{Cl} := 
	\inf \left \{ H(\mu) \, | \, (T,\mu) \; 
							{\rm is \; an \; invariant \; HMM \; of  \;} 
							\overleftrightarrow{X} 
        \right \}.
$$
\end{Def}

L\"ohr showed that for every stationary stochastic process there exists an 
invariant HMM $(T,\mu)$ such that
$$
H(\mu) = C_{Cl}
$$
hold and the infimum in Definition \ref{gencomplexity} is actually a minimum 
(Corollary 4.14 in \cite{Loe10}). In the following we denote this invariant 
HMM as {\em minimal HMM}.

The generative complexity is an upper bound for the excess entropy \cite{Loe10}
\begin{equation} \label{ELessC}
E \le C_{Cl}.
\end{equation}

In this paper we only consider processes which can be modeled by a minimal HMM 
with finitely many internal states. Markov processes of finite order are 
examples for processes with a finite set of internal states. Assuming finitely 
many internal states $\kS = \{S_1, \ldots, S_n \}$, we can write the initial 
distribution as a probability vector $\mu := (p_i)_{i=1}^n$ and the generator 
as a set of substochastic $n \times n$ matrices $T^{(r)}$ with entries 
$T^{(r)}_{i,j} := T(S_i)(S_j,r)$ for all $r \in \Sigma$. Since we are 
considering only a finite set of internal states, $C_{Cl}$ is always finite 
and with (\ref{ELessC}) the excess entropy (\ref{ExcessDef}) is also finite.

\section{$\epsilon$-Machines and minimal HMMs} \label{sec:minhmm}

The following construction of a transition-emitting HMM is often regarded in 
the literature and the resulting HMM coincide in many cases with a minimal 
HMM. Unfortunately not in any case this construction leads to a minimal HMM as 
often wrongly claimed in the literature (see \cite{Loe10,Loe09b,Loe09c} for 
counterexamples).
On the set $\Sigma^{\setN}$ of all past trajectories of the process 
$\overleftrightarrow{X}$ we define an equivalence relation \cite{Sha01}
\begin{equation} \label{equrel}
x \sim x' : \iff 
	P(\overrightarrow{X} \in \overrightarrow{x}| \overleftarrow{X} = x) 
	= P(\overrightarrow{X} \in \overrightarrow{x}| \overleftarrow{X} = x'), 
	\quad \forall \,  \overrightarrow{x} \in \overrightarrow{\kC},
\end{equation}
where $x , x' \in \Sigma^{\setN}$, $\overrightarrow{\kC}$ is the product 
$\sigma$-algebra generated by cylinder sets on $\Sigma^\setN$ and 
$P(\overrightarrow{X} \in \overrightarrow{x} |\overleftarrow{X} = x)$ is a 
regular version\footnote{
$P(\overrightarrow{X} \in \overrightarrow{x} |\overleftarrow{X} = x)$ is 
called a regular version if it is a Markov kernel.} 
of the conditional expectation.
The equivalence classes 
$$
S(x) := \{x' \in \Sigma^{\setN} | x' \sim x\}
$$ 
of relation (\ref{equrel}) are called {\em causal states} and are the internal 
states of the constructed HMM. The set of all causal states is denoted by 
$\kS := \{S(x) | x \in \Sigma^{\setN} \}$ and is measurable (Lemma 3.18 in 
\cite{Loe10}). In general there can be uncountably many causal states 
\cite{Cru94,Loe10,Loe09b} and the causal states depend on the version of 
conditional probability used in the definition \cite{Loe10}. We say that the 
number of causal states is finite if there exists a version of conditional 
probability such that there are only finitely many equivalence classes. A 
characteristic property of causal states is that they induce a minimal 
sufficient memory\footnote{A {\em memory kernel} is a Markov kernel 
$\gamma:\Sigma^{\setN} \rightarrow \kP(\kS)$. The associated random variable 
$M$ is called {\em memory variable} or simply {\em memory}. A memory variable 
is called {\em sufficient} if 
$P(\overrightarrow{X} \in A, \overleftarrow{X} \in B | M)
	= P(\overrightarrow{X} \in A|M) P(\overleftarrow{X} \in B | M)$ 
a.s. for all measureable sets $A,B$. A memory is {\em minimal} if every other 
sufficient memory has at least the same number of internal states and the 
corresponding memory variable has at least the same entropy, Corollary 3.21 in 
\cite{Loe10}.}. 

We are only considering stationary stochastic processes with a finite set of 
causal states $\kS = \{S_1, \ldots, S_n\}$. Given a past observation of 
infinite length $x_{-\infty}^t \in \Sigma^{\setZ}$ at time $t \in \setZ$ using 
stationarity we identify this shifted past with a causal state 
$S(\sigma^{-t-1}(x_{-\infty}^t)) \in \kS$. Together with the next symbol 
$x_{t+1}$ generated by the process the next causal state 
$S(\sigma^{-t-2}(x_{-\infty}^t x_{t+1})) \in \kS$ is uniquely determined and 
the causal states are Markov \cite{Sha01, Loe10}. We define the Markov kernels 
between two causal states $S_i, S_j \in \kS$ emitting an output symbol 
$r \in \Sigma$ for any $t\in \setZ$ as follows
\begin{eqnarray*}
T_{i,j}^{(r)} & := & T(S_i)(S_j, r) \\
	& = & P \left ( S(\sigma^{-t-2}(x_{-\infty}^t x_{t+1})) = S_j \, {\rm and} \, 
				  X_{t+1} = r \, \left | \, S(\sigma^{-t-1}(x_{-\infty}^t)) = S_i 
									  \right. 
		     \right ).
\end{eqnarray*}

The probability of a causal state $S_i \in \kS$ is denoted by $p_i := P(S_i)$.
The ordered pair $(T, (p_1, \ldots, p_n))$ is called {\em $\epsilon$-machine}. 
The $\epsilon$-machine is a transition-emitting HMM and a model for the 
original stochastic process \cite{Loe10, Loe09a}. 

\begin{Bem}
In general the $\epsilon$-machine is not the HMM with minimal number of 
internal states and also not the one with minimal classical internal state 
entropy. To be precise L\"ohr proved in \cite{Loe10} that for a countable 
alphabet $\Sigma$ the $\epsilon$-machine is the minimal partially deterministic
 HMM\footnote{An invariant HMM $(T,\mu)$  with measureable spaces 
$(\Sigma, \kD)$ and $(\kS, \kG)$ is called {\em partially deterministic} if 
there is a measureable function $f:\kS \times \Sigma \rightarrow \kS$ 
(transition function), such that for $\mu$-almost all $s \in \kS$ we have 
$T(s)(G \times D) = K_s(D \cap f(s,.)^{-1}(G)) \qquad \forall D \in \kD, G \in \kG$, 
where $K_s (.) := T(s)(\kS \times .)$ is the output kernel.} of the process 
$\overleftrightarrow{X}$. 
\end{Bem}

The $\epsilon$-machine has classical internal state entropy
$$
C_{\epsilon} := H(\kS) = - \sum_{j=1}^n p_j \log p_j,
$$
which is also known as {\em statistical complexity} \cite{Gra86, Sha01}. Since 
the generative complexity is an upper bound for the excess entropy, the 
statistical complexity is also an upper bound for the excess entropy 
\cite{Sha01, Cru03}
\begin{equation} \label{complexcessineq}
E \le C_{\epsilon}.
\end{equation}

The next theorem gives a characterization when (\ref{complexcessineq}) is 
strict.

\begin{HSatz} \label{Theorem1}
Given a stationary stochastic process $\overleftrightarrow{X}$ with excess 
entropy $E$ and statistical complexity $C_{\epsilon}$. Let its corresponding 
$\epsilon$-machine have transition probabilities $T_{i,j}^{(r)}$. Then 
$C_{\epsilon} > E$ if and only if there exists a non-zero probability that two 
different causal states $S_j$ and $S_k$ will both make a transition to a 
coinciding causal state $S_l$ upon emission of a coinciding output 
$r \in \Sigma$, i.e. $T_{j,l}^{(r)}, T_{k,l}^{(r)} \neq 0$.
\end{HSatz}
\begin{Proof}
Theorem 1 in \cite{GuW11}.
\end{Proof}

As a next step we extend the last theorem from $\epsilon$-machines to minimal 
HMMs.
We want now return to the general case and consider minimal HMMs which we 
denote as minimal classical models. From the definitions of the internal 
entropies it is clear that
\begin{equation} \label{clexcessinequ}
E \le C_{Cl} \le C_{\epsilon}.
\end{equation}
There exists examples such that $C_{Cl} < C_{\epsilon}$ holds and it is known 
that \cite{Loe10}
$$
C_{Cl} < C_{\epsilon} \Rightarrow E < C_{Cl},
$$
or the negation of this
\begin{equation} \label{clexcessequal}
E = C_{Cl} \Rightarrow C_{Cl} = C_{\epsilon}.
\end{equation}

With this fact it is possible to generalize Theorem \ref{Theorem1}.

\begin{HSatz} \label{GenTheorem}
Given a stationary stochastic process $\overleftrightarrow{X}$ with excess 
entropy $E$ and generative complexity $C_{Cl}$. Let its corresponding minimal 
HMM have transition probabilities $T_{i,j}^{(r)}$. Then $C_{Cl} > E$ if and 
only if there exists a non-zero probability that two different internal states 
$S_j$ and $S_k$ will both make a transition to a coinciding internal state 
$S_l$ upon emission of a coinciding output $r \in \Sigma$, i.e. 
$T_{j,l}^{(r)}, T_{k,l}^{(r)} \neq 0$.
\end{HSatz}
\begin{Proof}
With (\ref{clexcessequal}) and (\ref{clexcessinequ}) we get 
$E = C_{Cl} \iff C_{Cl} = C_{\epsilon}$. With Theorem \ref{Theorem1} and the 
negation of the last expression we yield the result.
\end{Proof}

\begin{Bem}
Theorem \ref{GenTheorem} shows that there is a kind of redundance in the 
minimal HMM producing the gap between $E$ and $C_{Cl}$. This redundance is an 
indicator for a possible improvement of the classical minimal HMM, see 
Theorem \ref{ThmBetterQuantumExists}.
\end{Bem}

\section{Hidden Quantum Markov Models and Holevo-Bound} \label{sec:qua}

Based on the classical minimal HMM introduced in Section \ref{sec:pre} it is 
possible to define quantum models with the same statistical behaviour. In the 
spirit of classical HMM we define a quantum version of such models introduced 
in \cite{Mon11} to reproduce a given stochastic process.

\begin{Def}[\cite{Mon11}]
A {\bf quantum operation} 
$\kK_r: {\rm Mat}(d,\setC) \rightarrow {\rm Mat}(d,\setC)$ is a completely 
positive, trace non-increasing linear map on the space of complex 
$d\times d$-matrices ${\rm Mat}(d, \setC)$. A {\bf Hidden Quantum Markov Model 
(HQMM)} is a density matrix $\rho \in {\rm Mat}(d,\setC)$ together with a set 
of quantum operations $\kK_r$, $\forall \,r \in \Sigma$ such that 
$\sum_{r \in \Sigma} \kK_r$ is trace-preserving.  At every time step a symbol 
$r \in \Sigma$ is generated with probability $P(r) = {\rm Tr}(\kK_r \rho)$ and 
the state vector is updated to $\rho_r = \kK_r \rho /P(r)$.
\end{Def}

There is an analogy between classical HMM and HQMM, for example the quantum 
operation $\kK_r$ plays the role of a substochastic matrix $T^{(r)}$ and the 
density matrix corresponds to the probability vector $(p_1, \ldots ,p_n)$, see 
\cite{Mon11} for more details. Furthermore it can be proved that for every 
transition-emitting HMM it is possible to construct a HQMM with the same 
statistical behaviour, i.e. the HQMM generates the same stochastic process 
\cite{Mon11}. This constructed HQMM is in general not unique and there are many
 possibilities to construct a HQMM producing the same stochastic process. In 
this paper we only consider constructions of HQMMs based on a given classical 
HMM. Before we write down such an explicit construction we will formulate the 
HQMM in the setting of a quantum channel. For this we introduce the general 
setting of a quantum channel.

Consider a finite {\em input alphabet} $\kX$ and a finite {\em output alphabet}
 $\kY$. Further let $\kH$ and $\kJ$ be the input and output Hilbert spaces. We 
want to transmit classical input data via a {\em quantum channel} that is, a 
completely positive, trace preserving map $\kE : B(\kH) \rightarrow B(\kJ)$, 
where $B(\kH)$ is the algebra of bounded operators acting on $\kH$. In order to
 do this choose an input random variable $X$ with values in $\kX$ and with a 
corresponding distribution $p:\kX \rightarrow [0,1]$. Code each $x \in \kX$ in 
a quantum state $\rho_x \in B(\kH)$ and after sending this through a quantum 
channel one can measure the output quantum state to get classical data as 
output. For every $y \in \kY$ there is a completely positive operator 
$\kK_y \in B(\kJ)$ such that $\sum_{y \in \kY} \kK_y = I_\kJ$, where $I_\kJ$ 
denotes the identity operator on $\kJ$. With ${\rm Tr_\kJ}$ we denote the 
partial trace with respect to $\kJ$. The probability that $y \in \kY$ is the 
output symbol, given $x \in \kX$ as input is
$$
T_{y,x} := {\rm Tr_\kJ} (\kE(\rho_x) \kK_y),
$$
and the output distribution takes the form
$$
\tilde p_y := 
	\sum_{x \in \kX} {\rm Tr_\kJ} (p_x \kE(\rho_x) \kK_y), 
				\qquad {\rm for \; every \;}y \in \kY.
$$
The corresponding random variable with distribution $(\tilde p_y)_{y \in \kY}$ 
and values in $\kY$ is denoted by $Y$.\\

We now give an explicit construction of a HQMM given a HMM which was defined 
in \cite{GuW11}. Without loss of generality let the finite alphabet be defined 
as $\Sigma := \{1, \ldots, M\}$. Given a classical HMM 
$\left (T, (p_1, \ldots, p_n) \right )$, with internal states 
$\kS = \{S_1, \ldots, S_n\}$. Choose as an input alphabet 
$\kX := \{1, \ldots, n\}$ and an output alphabet 
$\kY := \{1, \ldots, n\} \times \Sigma$. The Hilbert space takes the form 
$\kH := \setC^{n M} = \kJ$ and the quantum channel is defined as the identity 
$\kE:= Id_{B(\kH)}$. We code every $i \in \kX$ with {\em quantum internal 
states} as follows

\begin{equation} \label{quantummodel}
|S_i\rangle := 
	\sum_{r \in \Sigma} \sum_{j =1}^n \sqrt{T_{i,j}^{(r)}} |j \rangle \otimes |
						r \rangle \in \kH, \qquad \forall \, i \in \{1, \ldots,n \}.
\end{equation}

The corresponding density matrix is defined as 
$\rho_i := |S_i \rangle \langle S_i |$.
The HQMM takes the form $\rho := \sum_{i=1}^n p_i \rho_i$ and is equipped with 
quantum operations $\kK_{r,j} := P_{|j \rangle \otimes |r\rangle}$ which are 
projections on the space spanned by $|j \rangle \otimes |r \rangle$. Clearly 
$\sum_{r \in \Sigma, j\in \{1, \ldots, n\}} \kK_{r,j}$ is trace-preserving.
Consider $|S_i \rangle$ as an initial quantum internal state then with the 
projections $\kK_{r,j}$ it follows that
$$
T_{jr,i} = {\rm Tr} (\kE(\rho_i) \kK_{r,j}) = T_{i,j}^{(r)},
$$
holds. We set $x_0 = r$ as output and prepare the next quantum internal state 
$|S_j \rangle$. Repeating this procedure we get a sequence of symbols 
$x_0, x_1, \ldots$ with the same probability as produced with the classical HMM
 initialized in a state $S_i$. 
This proves that this HQMM have the same statistical behaviour as the classical
 HMM, which means that boths models have the same transition probabilities 
between equivalent states.

\begin{Bem}
In \cite{GuW11} this construction is applied to classical $\epsilon$-machines 
and the HQMM is called {\em quantum $\epsilon$-machine}. Since we are 
considering minimal HMMs which need not to be $\epsilon$-machines we call the 
defined HQMM a {\em quantum model induced by a minimal HMM}.
\end{Bem}

The quantum internal state entropy of a HQMM is the von Neumann entropy 
$$
C_q := S(\rho) := -{\rm Tr} \rho \log \rho.
$$
$C_q$ is the quantum version of the classical internal state entropy $H(\mu)$ 
and is bounded by this internal state entropy and especially by the generative 
complexity $C_{Cl}$.
\begin{Satz} \label{Upperbound}
Suppose $\rho = \sum_{j=1}^n p_j \rho_j$ where $p=(p_j)_{j=1}^n$ is a 
probability vector with $\sum_{j=1}^n p_j =1$ and the 
$\rho_j := |S_j\rangle \langle S_j|$ are density operators for every 
$j \in \{1,\ldots, n\}$. Then
$$
C_q \le  H \left ( p \right ),
$$
with equality if and only if the quantum internal states $|S_j \rangle$ are 
mutually orthogonal.
In especially given a minimal HMM $(T,p)$ with an induced quantum model 
$\rho$ we have
$$
C_q \le C_{Cl}.
$$
\end{Satz}
\begin{Proof}
Theorem 11.10 in \cite{Nie00} or alternatively an adaption of Theorem 3.7 in 
\cite{Pet08}.
\end{Proof}

\begin{Bem} \label{bem:quaprop}
In the case that the classical minimal HMM coincide with the classical 
$\epsilon$-machine it is not clear if the quantum internal states of the 
induced quantum model share the same properties as the classical causal 
states, i.e. the question if quantum internal states are minimal sufficient in 
the sense of quantum mechanics is not yet answered.
\end{Bem}

The next proposition is the well-known Holevo-Bound and gives an upper bound 
for the mutual information between classical input and classical output data.
\begin{Satz}[Holevo-Bound] \label{Holevobound}
Given the setting above with classical input random variable $X$ and classical 
output random variable $Y$, the following bound holds
\begin{equation} \label{holevoinequality}
I(X; Y) \le S(\rho) - \sum_{i=1}^n p_i S(\rho_i),
\end{equation}
where $\rho = \sum_{i=1}^n p_i \rho_i$ and with equality if and only if all 
$\rho_i$ commute.
\end{Satz}
\begin{Proof}
Theorem 12.1 in \cite{Nie00} or Theorem 7.3 in \cite{Pet08}. For the equality 
condition see for example \cite{Rus02}.
\end{Proof}

In the case that the HMM is an $\epsilon$-machine the lefthand side of 
(\ref{holevoinequality}) is the excess entropy.

\begin{Satz} \label{PropExcessCalc}
Let $(T,(p_1, \ldots, p_n))$ be an $\epsilon$-machine then given the setting 
above it holds that
$$
I(X;Y) = I(\overleftarrow{X}; \overrightarrow{X}) = E.
$$
\end{Satz}
\begin{Proof}
To prove the proposition we use a four variable mutual information introduced 
in \cite{Yeu91} and follow the same strategy as in \cite{Cru10}. For random 
variables $X, Y, Z, U$ we define
\begin{eqnarray*}
I(X; Y; Z; U) & := & I(X;Y;Z) - I(X; Y; Z |U), \\
I(X;Y;Z)      & := & I(X;Y) - I(X;Y|Z), \\
              &    & {\rm with \;} I(X;Y|Z) := H(X|Z) - H(X|Y,Z), \\
I(X; Y; Z |U) & := & I(X;Y|U) - I(X;Y|Z;U), \\
              &    & {\rm with \;} I(X;Y|Z;U) := H(X|Z,U) - H(X |Z,U,Y).
\end{eqnarray*}
Furthermore we use the following two identities which hold for a measurable 
function $f$ of a random-variable $X$ (\cite{Gra11}, Lemma 3.12)
\begin{equation} \label{ProofhelpEquations}
H(f(X)|X) = 0, \qquad H(X, f(X)) = H(X).
\end{equation}
We define mappings $g: \Sigma^\setN \rightarrow \kX$, with $g(\sigma) := j$ if 
$\sigma \in S_j$ and $f:\Sigma^{-\setN_0} \rightarrow \kY$, with 
$f(\sigma \sigma_0) := (i, \sigma_0)$ if $\sigma \in S_i$. Since we are 
considering $\epsilon$-machines $g$ and $f$ are well-defined and measurable. 
Thus we can write $X = g(\overrightarrow{X}), Y = f(\overleftarrow{X})$ and 
using (\ref{ProofhelpEquations}) we get
\begin{eqnarray}
H(Y | \overleftarrow{X}) & = & 0, \qquad \qquad \qquad \, \, 
			H(X | \overrightarrow{X}) = 0,  \label{condEntrFuncDisappear} \\
H(\overleftarrow{X}, Y ) & = & H(\overleftarrow{X}), \qquad \qquad 
			H(\overrightarrow{X}, X) = H(\overrightarrow{X}), \\ 
H(\overrightarrow{X}| \overleftarrow{X}, Y) & = & H(\overrightarrow{X} |Y), 
			\qquad \; H(\overleftarrow{X} | \overrightarrow{X}, X) = 
							H(\overleftarrow{X} | X). \label{condEntropyDisappear}
\end{eqnarray}
In the next step we show 
$I(\overrightarrow{X}; \overleftarrow{X}; X; Y) = 
	I(\overrightarrow{X}; \overleftarrow{X}) = E$. Consider
\begin{equation} \label{secondfourMITerm}
I(\overrightarrow{X}; \overleftarrow{X} ; X | Y) = 
	I(\overrightarrow{X} ; \overleftarrow{X} | Y) - 
		I(\overrightarrow{X} ; \overleftarrow{X} | X ; Y),
\end{equation}
then the first term disappear because with (\ref{condEntropyDisappear}) it 
holds
$$
I(\overrightarrow{X} ; \overleftarrow{X} | Y) = 
	H(\overrightarrow{X} |Y) - H(\overrightarrow{X} | \overleftarrow{X} , Y) 
	\stackrel{(\ref{condEntropyDisappear})}{=} 0.
$$
The second term of (\ref{secondfourMITerm}) is also zero, since
$$
I(\overrightarrow{X}; \overleftarrow{X} | X ; Y) = 
	H(\overrightarrow{X} |X , Y) - 
		H(\overrightarrow{X} | X, Y , \overleftarrow{X}) 
	\stackrel{(\ref{condEntropyDisappear})}{=} 0.
$$
Putting all together we yield
$$
I(\overrightarrow{X}; \overleftarrow{X} ; X | Y) = 0.
$$
Furthermore we have
$$
I(\overrightarrow{X}; \overleftarrow{X}; X) = 
	I(\overrightarrow{X}; \overleftarrow{X}) - 
		I(\overrightarrow{X}; \overleftarrow{X} |X) = 
	I(\overrightarrow{X}; \overleftarrow{X}),
$$
since $I(\overrightarrow{X} ; \overleftarrow{X} |X) = 
	H(\overleftarrow{X} | X) - H(\overleftarrow{X} | \overrightarrow{X} , X) 
	\stackrel{(\ref{condEntropyDisappear})}{=} 0$.
Finally we get
$$
I(\overrightarrow{X}; \overleftarrow{X}; X; Y) = 
	I(\overrightarrow{X}; \overleftarrow{X}).
$$
In a second step we show 
$I(\overrightarrow{X}; \overleftarrow{X}; X; Y) = I(X; Y)$. As in the first 
step the following term vanish
\begin{equation} \label{secondfourMITerm2}
I(X; Y ; \overrightarrow{X} | \overleftarrow{X}) = 
	I(X ; Y | \overleftarrow{X}) - 
		I(X; Y | \overrightarrow{X}; \overleftarrow{X}) = 0,
\end{equation}
since $I(X; Y |\overleftarrow{X} ) = 
	H(Y|\overleftarrow{X}) - H(Y| X , \overleftarrow{X}) 
	\stackrel{(\ref{condEntrFuncDisappear})}{=} 0$ and
$$
I(X; Y | \overrightarrow{X}; \overleftarrow{X}) = 
	H(X | \overrightarrow{X}, \overleftarrow{X}) - 
		H(X | Y , \overrightarrow{X}, \overleftarrow{X}) 
	\stackrel{(\ref{condEntrFuncDisappear})}{=} 0.
$$
Consider now
$$
I(X; Y; \overrightarrow{X}) = I(X; Y) - I(X; Y|\overrightarrow{X}),
$$
then the second term disappear, since
$$
I(X; Y | \overrightarrow{X}) = 
	H(X | \overrightarrow{X}) - H(X |Y,\overrightarrow{X}) 
	\stackrel{(\ref{condEntrFuncDisappear})}{=} 0.
$$
Thus we yield
$$
I(\overrightarrow{X}; \overleftarrow{X}; X; Y) = I(X; Y),
$$
and finally we get
$$
E = I(\overrightarrow{X}; \overleftarrow{X} ) = I(X; Y).
$$
\end{Proof}

The converse of Proposition \ref{PropExcessCalc} is not true as can be seen in 
the example treated in Section \ref{sec:ex}.

\begin{Bem} \label{RemExcessCalc}
In general it is difficult to calculate the excess entropy of a given 
stationary stochastic process. If one has given an $\epsilon$-machine for a 
process it is easy to calculate $I(X;Y)$ which coincide with the excess 
entropy $E$. Compared to the method in \cite{Ell09} which uses the structure 
of the $\epsilon$-machine, this is an alternative method to calculate $E$.
\end{Bem}

Since $I(X;Y)$ depends on the classical HMM we sometimes write $I_{HMM}(X;Y)$ 
if a distinction is necessary.
For general HMMs and especially for minimal HMMs which are not an 
$\epsilon$-machine the excess entropy is in general smaller than $I(X;Y)$ as 
the next example shows. This example can be found in \cite{Loe09c}.
 
\begin{Bsp} \label{counterex}
Let $\Sigma := \{0,1\}$ and consider a stationary Markov process generated by 
the $\epsilon$-machine $(T,(p_0, p_1))$ with $p_0 = p_1 = \frac{1}{2}$ and 
$$
T^{(0)} = \left ( \begin{array}{c c} 
							\frac{1}{2}(1+\epsilon) & 0 \\ 
							\frac{1}{2}(1-\epsilon) & 0 
						\end{array} 
			 \right ), \qquad 
T^{(1)} = \left ( \begin{array}{c c} 
							0 & \frac{1}{2}(1-\epsilon) \\ 
							0 & \frac{1}{2}(1+\epsilon) 
						\end{array} 
			 \right ),
$$
where $0 < \epsilon \le 1$. The statistical complexity is 
$C_\epsilon = 1$ for $\epsilon > 0$ and the excess entropy amounts to
$$
E = \frac{1}{2} 
			\left ( 
					( 1+\epsilon)\log(1+\epsilon) + ( 1-\epsilon)\log(1-\epsilon) 
			\right ),
$$
and coincide with $I_{Markov}(X;Y)$.
We give now a HMM which generates the same process (see \cite{Loe09c}), but 
with three internal states and smaller internal state entropy than 
$C_\epsilon$. Let $\kS := \{0, 1, 2\}$ with 
$$
T^{(0)} = \left ( \begin{array}{c c c} 
							\epsilon & 0 & 1-\epsilon \\ 
							0 & 0 & 0 \\ 
							\frac{\epsilon}{2} & 0 & \frac{1-\epsilon}{2} 
						\end{array} 
			 \right ), \qquad 
T^{(1)} = \left ( \begin{array}{c c c} 
							0 & 0 & 0 \\ 
							0 & \epsilon & 1-\epsilon \\ 
							0 & \frac{\epsilon}{2} & \frac{1-\epsilon}{2} 
						\end{array} 
			 \right ),
$$
and initial distribution $(p_0, p_1, p_2)$
$$
p_i = \left \{ \begin{array}{l l} 
						\frac{\epsilon}{2}, & {\rm if} \; i \in \{0,1\} \\ 
						1-\epsilon, & {\rm if} \; i = 2 
					\end{array} 
		\right..
$$
The internal state entropy of this HMM is given by
$$
H(p) = -(1-\epsilon)\log(1-\epsilon) - \epsilon \log \left (
																			\frac{\epsilon}{2} 
																	  \right).
$$ 
It is easy to calculate the lefthand side of the Holevo-Bound
$$
I_{3state}(X; Y) = \epsilon.
$$
For $\epsilon \in (0,1)$ the excess entropy is always strictly smaller than 
$I_{3state}(X;Y)$. Especially for $\epsilon$ small enough the three state HMM 
has smaller internal state entropy $H(p)$ than the $\epsilon$-machine as can 
be seen in Figure \ref{fig:counterexample}.

\begin{figure}[h]
\begin{center}
\epsfig{figure=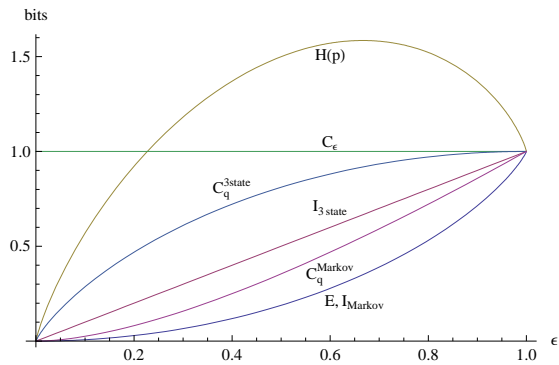,scale=0.85}
\caption{Excess entropy $E$, $I_{3state}:=I_{3state}(X;Y)$, $C^{3state}_q$, 
$I_{Markov} :=I_{Markov}(X;Y)$, $C^{Markov}_q$, internal state entropy $H(p)$ 
of the three state HMM described in Example \ref{counterex} and statistical 
complexity of the $\epsilon$-machine.}
\phantomsection \label{fig:counterexample}
\end{center}
\end{figure}
Furthermore L\"ohr showed in \cite{Loe09c} that the internal state entropy of 
the minimal HMM is bounded from below by
$$
C_{Cl} \ge -(1-\epsilon/2)\log(1-\epsilon/2)-\epsilon/2\log(\epsilon/2),
$$
where the lower bound coincide with the internal state entropy $C^{3state}_q$ 
of the quantum model induced by the three state HMM. 
This example shows that it is possible that the excess entropy is smaller than 
the lower-bound $I_{3state}(X;Y)$ of $C_q^{3state}$ given by the Holevo-Bound. 
Furthermore it also shows that even if the three state HMM has smaller 
internal entropy for sufficient small $\epsilon$, the internal state entropy 
$C_q^{Markov}$ of the quantum model induced by the markov model is strictly 
smaller than $C^{3state}_q$ and especially smaller than $I_{3state}(X;Y)$, see 
Figure \ref{fig:counterexample}. So it is not clear at all how minimal 
classical models and minimal quantum models are related to each other.
\end{Bsp}

Since the states $\rho_i = | S_i \rangle \langle S_i |$ are pure, we have 
$S(\rho_i) = 0$ so that Proposition \ref{Upperbound} and Proposition 
\ref{Holevobound} imply that in general
\begin{equation} \label{ECQC}
E \le I(X;Y) \le C_q \le C_{Cl},
\end{equation}
holds. 

\begin{Bem}
Inequality (\ref{ECQC}) allows us to compare the information stored in a 
classical minimal HMM and a induced quantum model which generate the same 
stochastic process. In order to compare the quantum internal state entropies 
of different HQMM constructions with the  internal state entropy of a given 
classical minimal HMM we have to ensure that (\ref{ECQC}) hold. Considering 
the right hand side of (\ref{holevoinequality}) the second term has to vanish 
and the internal states of such a HQMM has to fulfill 
$S(\rho_i) = 0, \quad \forall \, i \in \{1, \ldots, n\}$.
\end{Bem}

Gu et al. proved in \cite{GuW11} a remarkable theorem for classical 
$\epsilon$-machines that shows that if $C_{\epsilon} > E$ holds then the 
induced quantum model (\ref{quantummodel}) has internal state entropy strictly 
smaller than the internal state entropy of the classical $\epsilon$-machine 
$C_q < C_{\epsilon}$. We extend this result to classical minimal HMMs.

\begin{HSatz} \label{ThmBetterQuantumExists}
Given a stationary stochastic process $\overleftrightarrow{X}$ with excess 
entropy $E$ and generative complexity $C_{Cl}$ and $C_{Cl} > E$. Then there 
exists a quantum system that exhibits identical statistics with internal state 
entropy $C_q < C_{Cl}$.
\end{HSatz}
\begin{Proof}
Use Theorem \ref{GenTheorem} instead of Theorem \ref{Theorem1} in the proof of 
Theorem 2 in \cite{GuW11}.
\end{Proof}

In the next section we investigate equality conditions for these different 
internal state entropies.

\section{Equality conditions} \label{sect:equcond}

The next two propositions deliver a characterization when equality in the last 
two inequalities of (\ref{ECQC}) holds. 
\begin{Satz} \label{EqualTheorem}
Given a stationary stochastic process $\overleftrightarrow{X}$ with excess 
entropy $E$ and generative complexity $C_{Cl}$. Let the corresponding induced 
quantum model defined in (\ref{quantummodel}) have quantum internal state 
entropy $C_q$. Then it holds that
$E = I(X;Y) = C_q = C_{Cl}$ if and only if all quantum internal states are 
mutually orthogonal. 
\end{Satz}
\begin{Proof}
"$\Rightarrow$": It holds that $E = I(X;Y) = C_q = C_{Cl}$. Theorem 
\ref{GenTheorem} gives us that for each output $r \in \Sigma$, each index 
$l \in \{1, \ldots, n\}$ and each pair of indices $j \neq k$ it holds that one 
of the transition probabilities $T_{j,l}^{(r)}, T_{k,l}^{(r)}$ is zero. With 
the definition of the quantum internal states (\ref{quantummodel}) this 
implies $\langle S_j | S_k \rangle = 0$ for all indices $j \neq k$. \\
"$\Leftarrow$": The definition of the scalar product and 
$\langle S_j | S_k \rangle = 0$ for all indices $j \neq k$ imply that one of 
$T_{j,l}^{(r)}, T_{k,l}^{(r)}$ is zero for each output $r \in \Sigma$, index 
$l \in \{1, \ldots, n\}$ and pair of indices $j \neq k$. Again with Theorem 
\ref{GenTheorem} we get $E = C_{Cl}$. Together with (\ref{ECQC}) it follows 
that $E = I(X;Y) = C_q = C_{Cl}$.
\end{Proof}

\begin{Satz} \label{RedundanceTheorem}
Given a stationary stochastic process $\overleftrightarrow{X}$ with excess 
entropy $E$. For a given classical HMM generating $\overleftrightarrow{X}$ with
 internal state entropy $H(\mu)$ let the corresponding induced quantum model 
defined in (\ref{quantummodel}) have quantum internal state entropy $C_q$. Then
 it holds that
$E \le I(X;Y) = C_q < H(\mu)$ if and only if there exist at least two quantum 
internal states which are identical and all other quantum internal states are 
mutually orthogonal or also identical (i.e. 
$\exists \, k \neq i: \langle S_k | S_i \rangle = 1$, 
$\langle S_l | S_j \rangle $ 
is $0$ or $1$ for all other indices $l\neq j$).
\end{Satz}
\begin{Proof}
"$\Rightarrow$": Since $C_q < H(\mu)$ it follows from Proposition 
\ref{Upperbound} that not all quantum internal states are mutually orthogonal, 
i.e. there exist at least one pair of indices $i \neq k$ such that 
$\langle S_i | S_k \rangle \neq 0$. Furthermore Proposition \ref{Holevobound} 
implies that $I(X;Y) = C_q$ if and only if all density operators 
$\rho_i = |S_i \rangle \langle S_i|$ commute. It is easy to prove that all 
$\rho_i$ commute if and only if $\langle S_i | S_k \rangle = 0$ or 
$\langle S_i |S_k \rangle =1$ for all indices $i,k \in \{1, \ldots, n\}$. From 
this equivalence relation the claim follows. \\
"$\Leftarrow$": There exist at least one pair of indices $i \neq k$ such that 
$\langle S_i | S_k \rangle = 1$. Together with the definition of quantum 
internal states there is an $r \in \Sigma$ and an index 
$l \in \{1, \ldots, n\}$ such that $T^{(r)}_{k,l} \neq 0$ and 
$T^{(r)}_{i,l} \neq 0$. Since not all quantum internal states are mutually 
orthogonal it follows from Proposition \ref{Upperbound} that $C_q < H(\mu)$. 
From the Holevo-Bound (Proposition \ref{Holevobound}) we know that 
$I(X;Y) \le C_q$ with equality if and only if all density operators 
$\rho_i = |S_i \rangle \langle S_i|$ commute which is again equivalent to the 
condition that $\langle S_i | S_k \rangle = 1$ or 
$\langle S_i | S_k \rangle = 0$ for all indices $i,k \in \{1, \ldots, n\}$. 
Hence $I(X;Y) = C_q$ follows.
\end{Proof}

A direct consequence of Proposition \ref{RedundanceTheorem} is that if 
$E \le I(X;Y) = C_q < H(\mu)$ there exist two identical quantum internal states
 $\langle S_i | = \langle S_k |$, $i \neq k$. This implies that for all 
$r \in \Sigma$ and all indices $l \in \{1, \ldots, n\}$ it holds that 
$T_{i,l}^{(r)} = T_{k,l}^{(r)}$. Which means that in the corresponding 
classical HMM there are two states which are redundant and can be merged to 
one state. This HMM is not a classical minimal HMM for the underlying 
stochastic process as the next proposition shows. 

\begin{Satz}
Given a stationary stochastic process $\overleftrightarrow{X}$ with excess 
entropy $E$. For a given classical HMM generating $\overleftrightarrow{X}$ with
 internal state entropy $H(\mu)$ let the corresponding induced quantum model 
defined in (\ref{quantummodel}) have quantum internal state entropy $C_q$. The 
classical HMM corresponding to the induced quantum model in the case 
$E \le I(X;Y) = C_q < H(\mu)$ is not a classical minimal HMM and therefore has 
not minimal classical internal state entropy.
\end{Satz}
\begin{Proof}
Suppose that the classical HMM corresponding to the induced quantum model is a 
minimal HMM (i.e. $H(\mu) = C_{Cl}$), then one can remove all redundant states 
in this classical HMM and in the resulting induced quantum model there remains 
only orthogonal quantum internal states. With Proposition \ref{EqualTheorem} 
we have $E = I(X;Y) = C_q = C_{Cl}$ and the reduced classical HMM is in fact 
the minimal HMM which is an ideal model. So the not reduced classical HMM 
cannot be the minimal HMM which is a contradiction to the assumption and the 
claim is proved.
\end{Proof}

The last proposition implies that the case $E \le I(X;Y) = C_q < C_{Cl}$ 
cannot exist.

\begin{Bem}
The case $E \le I(X;Y) < C_q = C_{Cl}$ does not exist. Suppose this case 
exists. Then Proposition \ref{Upperbound} would imply that all quantum 
internal states are mutually orthogonal and Proposition \ref{EqualTheorem} 
implies $E = I(X;Y) = C_q = C_{Cl}$ which is a contradiction to the assumption.
\end{Bem}

That is given a minimal classical HMM one is either in the case that the 
classical HMM is as good as the induced quantum model or the induced quantum 
model has a quantum internal state entropy $C_q$ strictly smaller than 
$C_{Cl}$ and strictly greater than $I(X;Y)$. We summarize the different cases:

\begin{itemize}
	\item [(i)] $E = I(X;Y) = C_q = C_{Cl} \iff$ the classical HMM and the 
induced quantum model are both optimal and all quantum internal states are 
mutually orthogonal.
	\item [(ii)] $E \le I(X;Y) = C_q < C_{Cl}$ is not possible.
	\item [(iii)] $E \le I(X;Y) = C_q < H(\mu) \iff$ the corresponding 
classical model contains redundant states and is not a minimal HMM and the 
induced quantum model contains only orthogonal or identical states but at 
least two identical states.
	\item [(iv)] $E \le I(X;Y) < C_q = C_{Cl}$ is not possible.
	\item [(v)] $E \le I(X;Y) < C_q < C_{Cl} \iff$ the classical HMM can be 
optimal and there exists quantum internal states which are not orthogonal and 
not identical.
\end{itemize}

So if one chooses an optimal classical HMM which is not an ideal classical 
model, there is always an induced quantum model which is nearer to an ideal 
model but never achieve such an ideal model.

\section{Calculation Example} \label{sec:ex}

\begin{figure}[h]
\begin{center}
\subfigure[]{\epsfig{figure=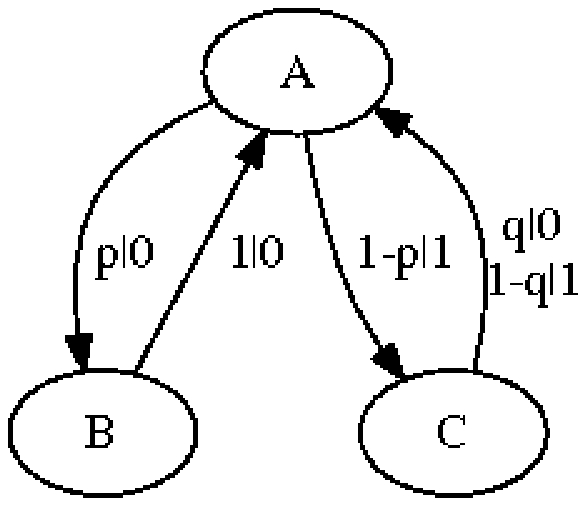,scale=0.6}} \qquad
\subfigure[]{\epsfig{figure=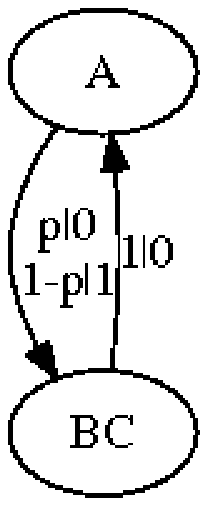,scale=0.6}}
\caption{(a) Minimal HMM for the RnC process. Nodes denoting the internal 
states of the HMM and edges labels $t | x$ give the probability 
$t=T_{S, S'}^{(x)}$ of making a transition from $S$ to $S'$ and seeing symbol 
$x$. (b) Minimal HMM for the underlying process in the case $q=1$.}
\phantomsection \label{fig:RnC}
\end{center}
\end{figure}

The following example illustrates the propositions shown in the preceding 
sections.
We consider the Random Noisy Copy HMM (RnC) \cite{Ell09}. This HMM generates a 
binary stochastic output process. It is given by a binary alphabet 
$\Sigma = \{0,1\}$, the internal states $\kS = \{A,B,C\}$ (which are also the 
causal states) and the Markov kernels 
$$
T^{(0)} = \left ( \begin{array}{c c c} 
							0 & p & 0 \\ 
							1 & 0 & 0 \\ 
							q & 0 & 0 
						\end{array} 
			 \right ), \qquad 
T^{(1)} = \left ( \begin{array}{c c c} 
							0 & 0 & 1-p \\ 
							0 & 0 & 0 \\ 
							1-q & 0 & 0 
						\end{array} 
			 \right ),
$$
with $0 \le p, q \le 1$. Figure \ref{fig:RnC} (a) shows a graphical 
representation of the RnC HMM.

The RnC HMM coincide with the classical $\epsilon$-machine. The left 
eigenvector of the stochastic matrix $T^{(0)} + T^{(1)}$ gives us the 
stationary distribution over the internal states
$$
P(\kS) = \frac{1}{2} \left ( \begin{array}{c c c} 
											1 & p & 1-p 
									  \end{array} 
							\right ).
$$
This allows us to calculate the generative complexity (which is identical 
with the statistical complexity)
$$
C_{Cl} = 1 + \frac{H(p)}{2},
$$
where $H(p) = -p\log(p) -(1-p)\log(1-p)$ is the binary entropy function. In 
this section logarithm is taken to the base $2$.
With more sophisticated techniques (see \cite{Ell09} for calculation details) 
or with Proposition \ref{RemExcessCalc} one can also calculate the excess 
entropy directly 
$$
E = I(X;Y) = 
	1 + \frac{H(p)}{2} - \frac{p+q(1-p)}{2}H	\left ( 
																\frac{p}{p + q(1-p)} 
															\right ).
$$

The quantum internal states defined in (\ref{quantummodel}) are
$$
|A \rangle = \left ( \begin{array}{c} 
								0 \\ 
								\sqrt{p} \\ 
								0 \\ 
								\sqrt{1-p} 
							\end{array} 
				 \right ), \qquad 
|B \rangle = \left ( \begin{array}{c} 
								1 \\ 
								0 \\ 
								0 \\ 
								0 
							\end{array} 
				 \right ), \qquad 
|C \rangle = \left ( \begin{array}{c} 
								\sqrt{q} \\ 
								0 \\ 
								\sqrt{1-q} \\ 
								0 
							\end{array} 
				 \right ).
$$

The eigenvalues of 
$\rho = 
	\frac{1}{2} \left ( |S_0 \rangle \langle S_0 | + 
									p |S_1 \rangle \langle S_1 | + 
									(1-p) |S_2 \rangle \langle S_2 | 
					\right )$ are 
$$
\left \{
	\frac{1}{2}, \frac{1}{4} \left 
										(1 \pm \sqrt{1 - 4p + 4p^2 + 4pq - 4p^2q} 
									 \right ) 
\right \}.
$$ 
Setting $\eta (x) := -x \log(x)$ the internal entropy of the induced quantum 
model amounts to
\begin{eqnarray*}
C_q = \eta \left ( \frac{1}{2} \right ) 
 & + & 	\eta \left (
					\frac{1}{4}\left (1 + \sqrt{1 - 4p + 4 p^2 + 4pq - 4p^2q} 
			 	 \right ) 
  			  \right ) \\
 & + & \eta \left (\frac{1}{4} \left (
											1 - \sqrt{1 - 4p + 4p^2 + 4pq - 4p^2q} 
										\right ) 
				\right ).
\end{eqnarray*}

\begin{figure}[h]
\begin{center}
\subfigure[q=0]{\epsfig{figure=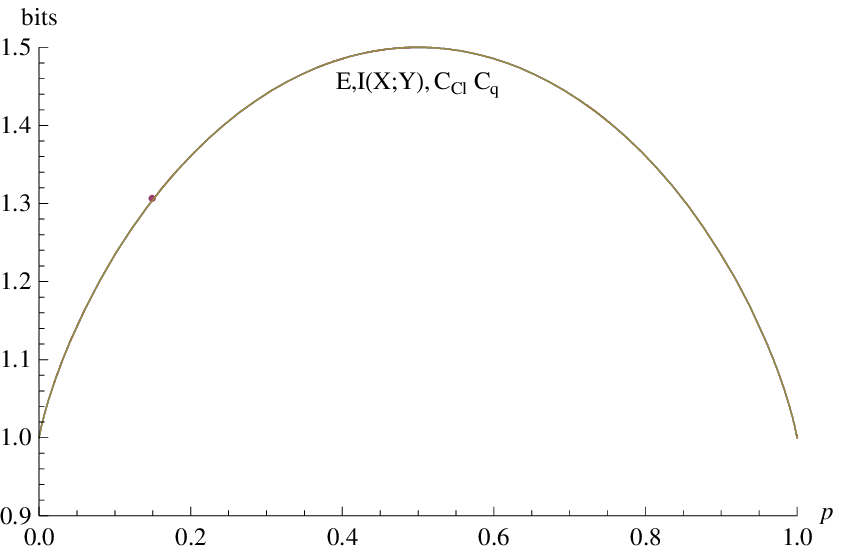,scale=0.54}} \quad
\subfigure[q=1]{\epsfig{figure=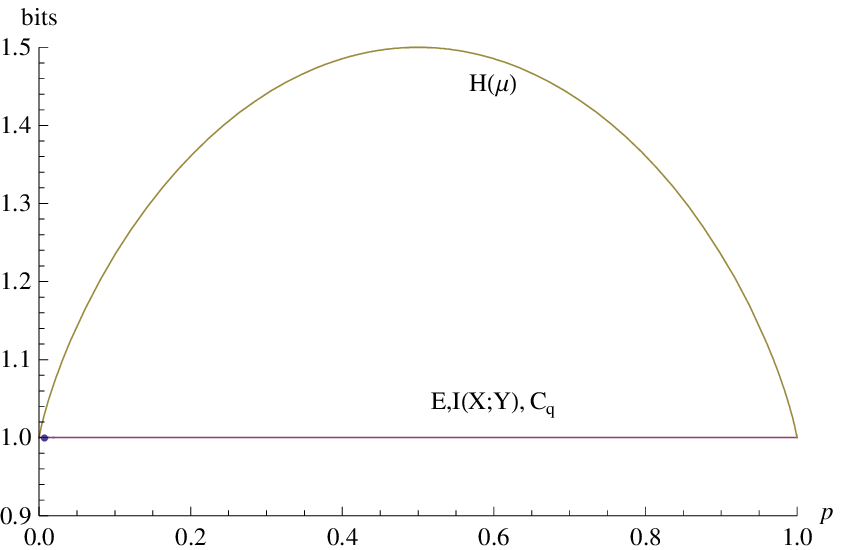,scale=0.54}} \quad
\subfigure[q=0.7]{\epsfig{figure=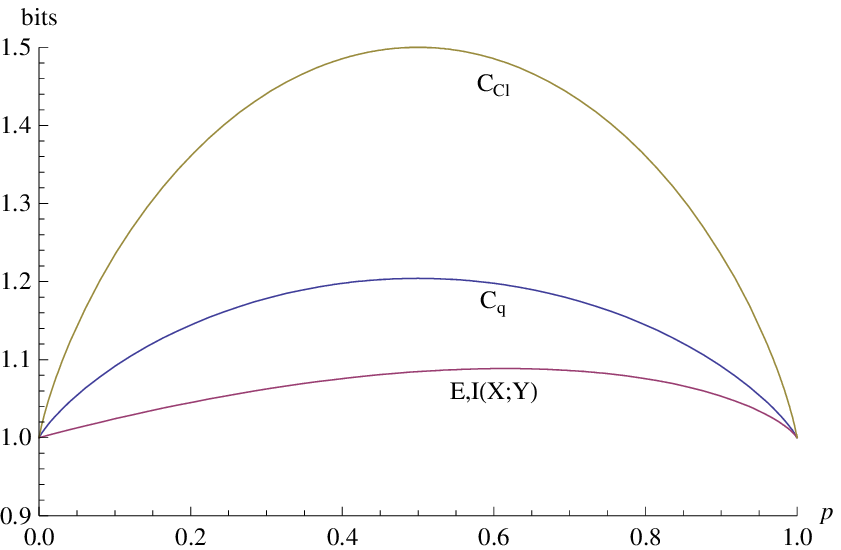,scale=0.54}}
\caption{Generative complexity $C_{Cl}$, quantum internal entropy $C_q$ and 
excess entropy $E = I(X;Y)$ for the RnC process with different $q$-values.}
\phantomsection \label{fig:RnCq}
\end{center}
\end{figure}

Fixing the parameter $q$ to certain values and varying $p$ we obtain the 
different cases described in Section \ref{sect:equcond}. 
For this we calculate the scalar product between the quantum internal states 
$\langle A | B \rangle = \langle A | C \rangle  = 0$ and 
$\langle B|C \rangle = \sqrt{q}$. 
Setting $q = 0$ all quantum internal states are mutually orthogonal and we are 
in case (i) which is shown in Figure \ref{fig:RnCq} (a). 

For $q=1$ the quantum internal states $|B\rangle$ and $|C\rangle$ are 
identical while $|A \rangle$ and $|B\rangle$ are orthogonal. Thus we are in 
case (iii) as seen in Figure \ref{fig:RnCq} (b). For $0 < q < 1$ we are in 
case (v) and have a gap between $E$, $C_q$ and $C_{Cl}$ as depicted in Figure 
\ref{fig:RnCq} (c) for $q=0.7$.

For $q=1$ the states $|B \rangle$ and $|C \rangle$ are identical and the 
corresponding classical HMM is not an $\epsilon$-machine but still $E=I(X;Y)$ 
holds for this model. This shows that the converse of Proposition 
\ref{PropExcessCalc} is not true. In the corresponding classical model (Fig. 
\ref{fig:RnC} (a)) the states $B$ and $C$ can merged to a state $BC$ (see Fig. 
\ref{fig:RnC} (b)). This is the classical minimal HMM for the underlying 
process.

\section{Alternative HQMMs} \label{sec:alternatives}

The induced quantum model (\ref{quantummodel}) introduced in Section 
\ref{sec:qua} is not the only possible HQMM construction that model a given 
stochastic process. In this section we present an alternative HQMM 
construction which is also able to model a stochastic process generated by a 
corresponding classical minimal HMM. For this we follow the construction 
suggested in \cite{Mon11}. Given a classical minimal HMM 
$\left (T, (p_1, \ldots, p_n ) \right )$ with internal states 
$\kS = \{S_1, \ldots, S_n\}$ we define internal states of the quantum model as 
$|i \rangle$ for $i \in \{1, \ldots,n \}$. Furthermore we have 
$\rho_i := |i \rangle \langle i |$ and define quantum operations with a sum 
representation\footnote{The Stinespring-Kraus Theorem shows that every 
completely positive map admits a (nonunique) operator-sum representation, so 
that can be written as $\kK \rho = \sum_i K_i \rho K_i^*$ where $K_i$ are 
linear operators on a Hilbert space, \cite{Kra83}.}
$$
\kK_r \rho := 
	\sum_{i,j=1}^n K_r^{i,j} \rho \left (K_r^{i,j} \right )^*, \qquad 
			K_r^{i,j} := \sqrt{T_{j,i}^{(r)}} |i \rangle \langle j |,
$$
for every symbol $r \in \Sigma$. With 
$\kK_r \rho_j = \sum_{i=1}^n T^{(r)}_{j,i} \rho_i$ we get
$$
P(X_0 = r|S_j) = {\rm Tr}( \kK_r \rho_j) = \sum_{i=1}^n T_{j,i}^{(r)} = 
			\sum_{i=1}^n P(X_0= r;S_i|S_j),
$$
and thus have the same transition probabilities as in the classical minimal 
HMM.

The quantum internal state entropy $\tilde C_q$ of this quantum model always 
coincide with the generative complexity of the process
$$
\tilde C_q = S\left ( \sum_{i=1}^n p_i \rho_i \right ) = 
				 H\left ( \{p_i\}_{i=1}^n \right) = C_{Cl}.
$$
In the next example treated in \cite{Mon11} we will see that in general 
$I(X;Y)$ is strictly smaller than $\tilde C_q$ and Proposition 
\ref{EqualTheorem} is not true for this type of HQMM construction. Consider 
the stochastic process generated by a classical 4-symbol HMM (which is minimal 
and coincide with the classical $\epsilon$-machine) with internal states 
$\kS = \{U,D,R,L\}$ and transition matrices

\begin{eqnarray}
T^{(0)} = \left ( \begin{array}{c c c c} 
							1/2 & 0 & 0 & 0 \\ 
							0 & 0 & 0 & 0 \\ 
							1/4 & 0 & 0 & 0 \\ 
							1/4 & 0 & 0 & 0
						\end{array} 
			 \right ), \qquad 
T^{(1)} = \left ( \begin{array}{c c c c} 
							0 & 0 & 0 & 0 \\ 
							0 & 1/2 & 0 & 0 \\ 
							0 & 1/4 & 0 & 0 \\ 
							0 & 1/4 & 0 & 0 
						\end{array} 
			 \right ), \nonumber \\
T^{(2)} = \left ( \begin{array}{c c c c} 
							0 & 0 & 1/4 & 0 \\ 
							0 & 0 & 1/4 & 0 \\ 
							0 & 0 & 1/2 & 0 \\ 
							0 & 0 & 0 & 0
						\end{array} 
			 \right ), \qquad 
T^{(3)} = \left ( \begin{array}{c c c c} 
							0 & 0 & 0 & 1/4 \\ 
							0 & 0 & 0 & 1/4 \\ 
							0 & 0 & 0 & 0 \\ 
							0 & 0 & 0 & 1/2 
						\end{array} 
			 \right ). \label{examplequ}
\end{eqnarray}

\begin{figure}[h]
\begin{center}
\epsfig{figure=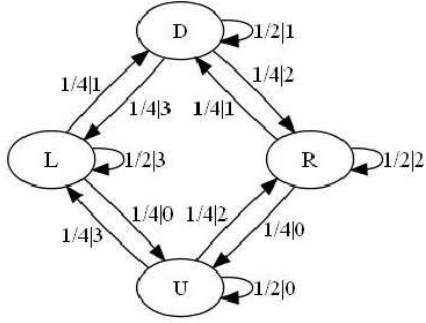,scale=0.6}
\caption{Classical 4-symbol HMM defined by equations (\ref{examplequ}).}
\phantomsection \label{fig:altexample}
\end{center}
\end{figure}

Figure \ref{fig:altexample} shows a graphical representation of this HMM.

We obtain as a stationary distribution
$$
P(\kS) = \frac{1}{4} \left ( \begin{array}{c c c c} 
											1 & 1 & 1 & 1 
									  \end{array} 
							\right ),
$$
and the generative complexity calculates to $C_{Cl} = 2$. With the framework 
introduced in Section \ref{sec:qua} it is possible to calculate $I(X;Y)$ which 
is the left hand side in (\ref{holevoinequality}) and amounts to 
$I(X;Y) = \frac{1}{2}$. Since $\tilde C_q = C_{Cl} = 2$ Proposition 
\ref{EqualTheorem} holds not in this situation. The quantum internal state 
entropy of the induced quantum model defined in Section \ref{sec:qua} is 
(logarithm is taken to the base $2$)
\begin{eqnarray*}
C_q & = & \frac{1}{8} \left ( \log(64) + \left (-3 + 2 \sqrt{2}\right ) 
													\log \left (
																\frac{1}{8} (3 - 2 \sqrt{2}) 
														  \right ) \right. \\
    & & \qquad \qquad \quad \left.
											 - \left (3 + 2 \sqrt{2} \right ) 
													\log \left (
																\frac{1}{8} (3 + 2 \sqrt{2})
														  \right ) 
							  \right ) \\
& \approx & 1.2018.
\end{eqnarray*}

Monras et al. suggest in \cite{Mon11} another quantum model for this process 
which is only a 2-level quantum system instead of the 4-level quantum system 
given above. Given the internal states 
$|\uparrow \rangle$, 
$\,|\downarrow \rangle$, 
$|+\rangle = \frac{|\uparrow \rangle + |\downarrow \rangle}{\sqrt{2}}$ 
and 
$|-\rangle = \frac{|\uparrow \rangle - |\downarrow \rangle}{\sqrt{2}}$ 
and quantum operations $\kK_r \rho = K_r \rho K_r^*$ for $r \in \{0,1,2,3\}$ 
with
$$
K_0 = \frac{1}{\sqrt{2}} |\uparrow \rangle \langle \uparrow |, \qquad 
K_2 = \frac{1}{\sqrt{2}} |+\rangle \langle +|,
$$
$$
K_1 = \frac{1}{\sqrt{2}} |\downarrow \rangle \langle \downarrow |, \qquad 
K_3 = \frac{1}{\sqrt{2}} |-\rangle \langle -|,
$$
it can be derived from this HQMM the same statistical behaviour as the 
classical HMM. The quantum internal state entropy of this quantum model is 
smaller than $C_q$ and amounts to
$$
S(\rho) = 1,
$$
with 
$\rho = \frac{1}{4} |\uparrow \rangle \langle \uparrow | \, + 
	  \, \frac{1}{4} |\downarrow \rangle \langle \downarrow | \, + 
	  \, \frac{1}{4} |+ \rangle \langle + | \, + 
	  \, \frac{1}{4} |- \rangle \langle - |$.

This example shows that in general the induced quantum model 
(\ref{quantummodel}) is not the one with minimum quantum internal state 
entropy. The structure of quantum models with minimal internal state entropy 
is an open question.
\\

\begin{Ack} 
I would like to thank Andreas Knauf for motivating me to work on this topic, 
for fruitful discussions and for suggestions to improve the text.
\end{Ack}


\begin{thebibliography}{6}
	\bibitem[Cov06]{Cov06}
		{\sc T. Cover and J. Thomas}:
		{\em Elements of Information theory}, 2nd ed.,
		{John Wiley \& Sons, Hoboken, New Jersey, 2006}.
	\bibitem[Cru83]{Cru83}
		{\sc J. Crutchfield and N.H. Packard}:
		{\em Symbolic dynamics of noisy chaos},
		{Phys. D\ {\bf 7}, 201 (1983)}.
	\bibitem[Cru94]{Cru94}
		{\sc J. Crutchfield}:
		{\em The calculi of emergence: Computation, dynamics and induction},
		{Phys. D\ {\bf 75}, 11 (1994)}.
	\bibitem[Cru03]{Cru03}
		{\sc J. Crutchfield and D. Feldman}:
		{\em Regularities Unseen, Randomness Observed: Levels of Entropy 
				Convergence},
		{Chaos\ {\bf 13}, 25 (2003)}.
	\bibitem[Cru10]{Cru10}
		{\sc J. Crutchfield and C. Ellison}:
		{\em The Past and the Future in the Present},
		{arXiv:1012.0356v1 (2010)}.
	\bibitem[Ell09]{Ell09}
		{\sc C. Ellison, J. Mahoney and J. Crutchfield}:
		{\em Prediction, Retrodiction and the amount of Information stored in the
			 Present},
		{J. Stat. Phys.\ {\bf 136}, 1005 (2009)}.
	\bibitem[Gra86]{Gra86}
		{\sc P. Grassberger}:
		{\em Toward a quantitative theory of self-generated complexity},
		{Internat. J. Theoret. Phys.\ {\bf 25}, 907 (1986)}.
	\bibitem[Gra11]{Gra11}
		{\sc R.M. Gray}:
		{\em Entropy and Information theory}, 2nd ed.,
		{Springer, New-York, 2011}.
	\bibitem[GuW11]{GuW11}
		{\sc M. Gu, K. Wiesner, E. Rieper and V. Vedral}:
		{\em Sharpening Occam's Razor with Quantum Mechanics},
		{arXiv:1102.1994v4 (2011)}.
	\bibitem[Kra83]{Kra83}
		{\sc K. Kraus}:
		{\em States, Effects and Operations},
		{Springer, Berlin, 1983}.
	\bibitem[Loe09a]{Loe09a}
		{\sc W. L\"ohr}:
		{\em Properties of the Statistical Complexity Functional and Partially 
			Deterministic HMMs},
		{Entropy\ {\bf 11}, 385 (2009)}.	
	\bibitem[Loe09b]{Loe09b}
		{\sc W. L\"ohr and N. Ay }:
		{\em On the Generative Nature of Prediction},
		{Adv. Complex Syst.\ {\bf 12}, 169 (2009)}.
	\bibitem[Loe09c]{Loe09c}
		{\sc W. L\"ohr and N. Ay}:
		{\em Non-Sufficient Memories that are Sufficient for Prediction},
		in Complex Sciences, 
		J. Zhou ed.,
		Springer, Berlin Heidelberg, 265 (2009).	
	\bibitem[Loe10]{Loe10}
		{\sc W. L\"ohr}:
		{\em Models of Discrete-Time Stochastic Processes and Associated 
		Complexity Measures},
		{PhD-Thesis, Leipzig, (2010)}.	
	\bibitem[Mon11]{Mon11}
		{\sc A. Monras, A. Beige and K. Wiesner}:
		{\em Hidden Quantum Markov Models and non-adaptive read-out of many-body 
		states},
		{Appl. Math. and Comp. Sciences\ {\bf 3}, 93 (2011)}.
	\bibitem[Nie00]{Nie00}
		{\sc M. Nielsen and I. Chuang }:
		{\em Quantum Computation and Quantum Information},
		{Cambridge University Press, 2000}.
	\bibitem[Pet08]{Pet08}
		{\sc D. Petz}:
		{\em Quantum Information Theory and Quantum Statistics},
		{Springer, Berlin Heidelberg, 2008}.
	\bibitem[Pin64]{Pin64}
		{\sc M.S. Pinsker}:
		{\em Information and Information Stability of Random Variables and 
		Processes},
		{Holden-Day, 1964}.
	\bibitem[Rus02]{Rus02}
		{\sc M. B. Ruskai}:
		{\em Inequalities for quantum entropy: A review with conditions for 
		equality},
		{J. Math. Phys.\ {\bf 43}, 4358 (2002)}.
	\bibitem[Sha01]{Sha01}
		{\sc C. R. Shalizi}:
		{\em Causal Architecture, Complexity and Self-Organization in Time},
		{PhD-Thesis, Madison, (2001)}.
	\bibitem[Yeu91]{Yeu91}
		{\sc R. Yeung}:
		{\em A New Outlook on Shannon's Information Measures},
		{IEEE Trans. Inform. Theory\ {\bf 37}, 466 (1991)}.
\end{thebibliography}
\end{document}